\documentclass{article}
\usepackage{natbib}
\usepackage{graphicx,color}

\begin{document}
\newcommand\affiliation[1]{\gdef\@affiliation{#1}}
\gdef\@affiliation{}
\newcommand\etal{\mbox{\textit{et al.}}}
\relax

\title{The Ionization State of the Halo Planetary Nebula NGC 2438\footnote{to appear in IAUS 283, proceedings of the IAU Symposium \protect{\newline}
\indent\ \ ``Planetary Nebulae: An Eye to the Future'', \protect{\newline}
\indent\ \ Eds.: A. Manchado, L. Stanghellini and D. Sch{\"o}nberner}}

\author
{Silvia Dalnodar}
\date{\scriptsize{
Institute of Astro- and Particle Physics, University Innsbruck, Technikerstrasse 25/8, \\A-6020 Innsbruck, Austria}}

\maketitle

\begin{abstract}
NGC 2438 is a classical multiple shell or halo planetary nebula (PN). Its central star and the main nebula are well studied. Also it was target of various hydrodynamic simulations (Corradi et al. 2000). This initiated a discussion whether the haloes are mainly containing recombined gas (Sch\"{o}nberner et al. 2002), or if they are still ionized (Armsdorfer et al. 2003).
An analysis of narrow-band images and long slit spectra at multiple slit positions was done to obtain a deeper look on morphological details and the properties of the outer shell and halo. For this work there was data available from ESO (direct imaging and long slit spectroscopy) and from SAAO (spectroscopic observations using a small slit - scanning over the whole nebula). Using temperature measurements from emission lines resulted in an electron temperature which clearly indicates a fully ionized stage. Additionally measurements of the electron density suggest a variation of the filling factor.

\noindent{{\it Keywords:} {\tt planetary nebulae: individual (NGC 2438)}}
\end{abstract}

\begin{figure}[ht]
\vspace{0.4 cm}
\begin{center}

\scalebox{0.45}{\includegraphics[angle=0]{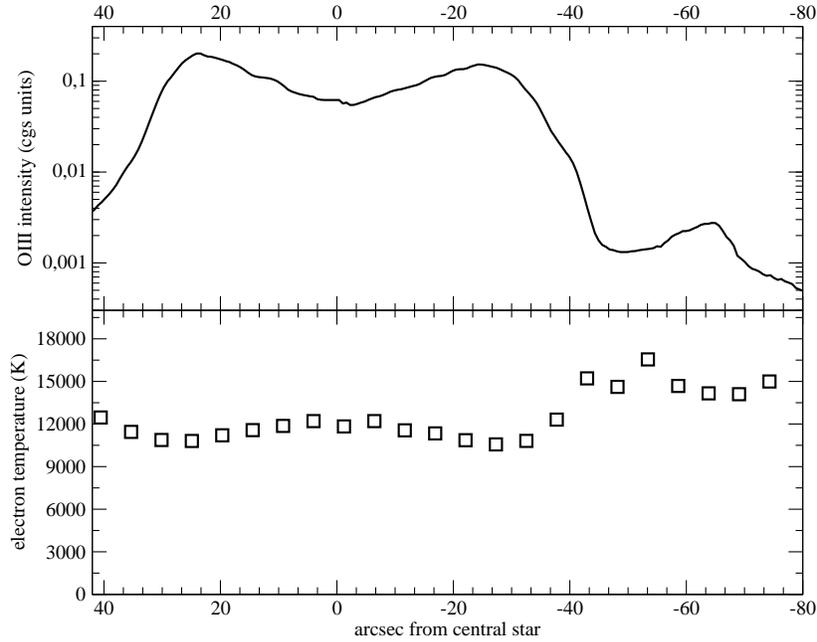}}

\vspace{0.2 cm}
\caption{This figure presents the electron temperature measurement resulting from the [OIII] lines. The upper part shows the [OIII] 5007 \AA\ logarithmic line intensity (line). The lower part shows the electron temperature profile over the main nebula and the western halo (boxes). The temperature is marginal higher in the inner part of the halo compared to the outer edge, where the temperature decreases again. To improve the S/N ratio in this evaluation, every time 5 arcseconds were integrated.
The temperature raises in thinner photoionized gaseous regions in the inner part of the halo and then slightly lowers at the outer edge of the halo. This is caused by a low filling factor. Due to this small filling factor the transmission of the radiation is even higher. This temperature measurements clearly imply that there can be no major recombination in this region.}
   \label{fig1}
\end{center}
\end{figure}
The data reduction was done using all the available data from ESO and from SAAO. The ESO-MIDAS astronomy data reduction package was used to process our obtained data. To confirm the quality of the results, comparisons between different nights and wavelengths were done. The variations are typically very small (just up to 10\%) and the overall quality of the flux calibration is very high.

The temperature in a nebula can be determined from measurements of ratios of intensities of particular pairs of emission lines. The essential line for electron temperature determination in PNe used here is the [OIII] line at 4363 \AA.

A calculation of the gradient of the logarithmic intensity of the different [OIII] emission lines (5007 \AA\ ,4958 \AA\ and 4363 \AA) was done. At the western part of the nebula the blue [OIII] line at 4363 \AA\ is also clearly visible in the halo (see figure 1). That allows for the first time the derivation of an electron temperature in the halo. The shape and gradient of the [OIII] lines are nearly parallel overall. This implies a rather homogeneous electron temperature.
The ratio of emission-line strengths is calculated by using the formulas described in Osterbrock \& Ferland (2006), where the coefficients result from numerical calculations of collision strengths and transition probabilities.

Using these temperature measurements from the emission lines result for the [OIII] lines in an electron temperature of about 10000 K to 13000 K in the main nebula. In the inner region of the halo the temperature raises up to about 15000 K to 17000 K and then slightly lowers again to a value of about 14000 K at the outer edge of the halo. This temperature measurements clearly indicate a fully ionized stage. No recombination can occur at such temperatures.

To verify this result, some CLOUDY (Ferland et al. 1998) calculations were done. With a temperature of 15000 K, more than 99\% of the Hydrogen is ionized. Looking at Helium, about 91\% of the atoms are in the HeII state. 94\% of the Oxygen atoms are in the OIII state. This confirms the result that there is hardly any recombination.

In this investigations the conclusion was made that the raising temperature in the inner part of the halo is caused by an even lower filling factor in the halo. This leads to a higher radiation energy and to a lower opacity. Recombination seems to be impossible in this stage. Also the results from the electron density measurements using the SAAO spectra suggest a variation of the filling factor.

\smallskip

\noindent{\sl Acknowledgements:\\}{\small S.D. is funded by Austrian Science Fund (FWF) DK+ project {\sl Computational Interdisciplinary Modeling}, W1227}
\vspace{5mm}

\centerline{\bf References:}
\noindent{Corradi, R.L.M., Sch\"{o}nberner, D., Steffen, M., \& Perinotto, M.} 2000, \textit{A\&A}, 354, 1071
\smallskip

\noindent{Sch\"{o}nberner, D., \& Steffen, M.} 2002, \textit{Rev. Mexicana AyA}, 12, 144
\smallskip

\noindent{Armsdorfer, B., Kimeswenger, S., \& Rauch, T.} 2003, \textit{Planetary Nebulae: Their Evolution and Role in the Universe}, Proc. IAU Symposium Vol. 209, 511
\smallskip

\noindent{Ferland, G.J., Korista, K.T., Verner,
D.A., Ferguson, J.W., Kingdon, J.B., \& Verner, E.M.}
1998, \textit{PASP}, 110, 761
\smallskip

\noindent{Osterbrock, D.E., \& Ferland, G.J.} 2006,
\textit{Astrophysics of gaseous nebulae and active galactic nuclei}, Sausalito, CA: University Science Books


\end{document}